\documentclass{article}
\usepackage{arxiv}
\usepackage[utf8]{inputenc}
\usepackage[english]{babel}
\usepackage{cite}
\usepackage{hyperref}
\usepackage{amssymb}
\usepackage{amsmath}
\usepackage{graphicx}
\usepackage{bm}
\usepackage{subcaption}
\usepackage{enumitem}

\newcommand{\phit}{\tilde{\phi}}
\newcommand{\psit}{\tilde{\psi}}
\newcommand{\xic}{x_\text{ic}^i}
\newcommand{\sic}{x_{0,\text{ic}}^i}
\newcommand{\xbc}{x_\text{bc}^i}
\newcommand{\tbc}{t_\text{bc}^i}
\newcommand{\sbc}{x_{0,\text{bc}}^i}
\newcommand{\srcf}{x_{0,f}^i}
\newcommand{\xf}{x_{f}^i}
\newcommand{\tf}{t_{f}^i}
\newcommand{\loss}{\mathcal{L}}

\title{Physics-informed neural networks for one-dimensional sound field predictions with parameterized sources and impedance boundaries}

\author{
  Nikolas Borrel-Jensen \\
  Acoustic Technology, Department of Electrical and Photonics Engineering\\
  Technical University of Denmark, 2800 Kongens Lyngby, Denmark\\
   \And
   Allan P. Engsig-Karup \\
   Department of Applied Mathematics and Computer Science\\ 
   Technical University of Denmark, 2800 Kongens Lyngby, Denmark \\  
  \AND
  Cheol-Ho Jeong \\
  Acoustic Technology, Department of Electrical and Photonics Engineering \\
  Technical University of Denmark, 2800 Kongens Lyngby, Denmark\\
}
\date{}

\begin{document}
\maketitle


\begin{abstract}
Realistic sound is essential in virtual environments, such as computer games and mixed reality. Efficient and accurate numerical methods for pre-calculating acoustics have been developed over the last decade; however, pre-calculating acoustics makes handling dynamic scenes with moving sources challenging, requiring intractable memory storage. A physics-informed neural network (PINN) method in 1D is presented, which learns a compact and efficient surrogate model with parameterized moving Gaussian sources and impedance boundaries, and satisfies a system of coupled equations. The model shows relative mean errors below 2\%/0.2 dB and proposes a first step in developing PINNs for realistic 3D scenes.
\end{abstract}

\section{Introduction}
In computer games and mixed reality, realistic sound is essential for an immersive user experience. The impulse responses (IR) can be obtained accurately and efficiently by numerically solving the wave equation using traditional numerical methods, such as finite element methods (FEM) \cite{Okuzono2014}, spectral element methods (SEM) \cite{Pind2019}, DG-FEM \cite{Melander:279868} and finite-difference time-domain methods (FDTD) \cite{Botteldoorena1995, Hamilton2017}. For real-time applications spanning a broad frequency range, the IRs are calculated offline due to the computational requirements. However, for dynamic, interactive scenes with numerous moving sources and receivers, the computation time and storage requirement for a lookup database become intractable (in the range of gigabytes) since the IR is calculated for each source-receiver pair. When covering the whole audible frequency range, these challenges become even more extensive. Previous attempts to overcome the storage requirements of the IRs include work for lossy compression \cite{Raghuvanshi2014}, and lately, a novel portal search method has been proposed as a drop-in solution to pre-computed IRs to adapt to flexible scenes, e.g., when doors and windows are opened and closed \cite{raghuvanshi2021dynamic}. A recent technique for handling parameter parameterization and model order reduction for acceleration of numerical models is the reduced basis method (RBM) \cite{Hesthaven2015, Llopis2021}. Although very efficient, RBM cannot meet the runtime requirements regarding computation time for virtual acoustics.

In this paper, we consider a new approach using physics-informed neural networks (PINNs)\cite{Psichogios1992, Lagaris1998, Raissi2019} including knowledge of the underlying physics (in contrary to traditional ``black box'' neural networks\cite{Fan2020}) to learn a surrogate model for a 1D domain that can be executed very efficiently at runtime (in the range of ms) and takes up little storage due to their intrinsic interpolation properties in grid-less domains. The applications of PINNs in virtual acoustics are very limited \cite{Moseley2020, Rasht-behesht}, and the main contribution of this work is the development of frequency-dependent and independent impedance boundary conditions with parameterized moving Gaussian sources, making it possible to model sound propagation taking boundary materials properly into account. This work investigated PINNs for virtual acoustics in a 1D domain -- still taking the necessary physics into account -- making it a possible stepping stone to model realistic and complex 3D scenes for applications, such as games and mixed reality, where the computation and storage requirements are very strict.

\section{Methods}
We take a data-free approach where only the underlying physics are included in the training and their residual minimized through the loss function, allowing insights into how well PINNs perform for predicting sound fields in acoustic conditions. The Gaussian impulse is used as the initial condition tested with frequency-independent and dependent impedance boundaries. To assess the quality of the developed PINN models, we have used our in-house open-source SEM simulators\cite{Pind2019}.

\subsection{Governing equations}
We consider in the following the use of PINNs for the construction of a surrogate model predicting the solution to the linear wave equation in 1D
\begin{equation}
  \frac{\partial^2 p(x,t)}{\partial t^2}-c^2\frac{\partial^2 p(x,t)}{\partial x^2}=0,\qquad t \in \mathbb{R^+}, \qquad x \in \mathbb{R}. \label{wave_equation}
\end{equation}
where $p$ is the pressure (Pa), $t$ is the time (s) and $c$ is the speed of sound in air (m/s). The initial conditions (ICs) are satisfied by using a Gaussian source for the pressure part and setting the velocity equal to zero
\begin{equation}
p(x,t=0,x_0) = \exp\left[-\left(\frac{x-x_0}{\sigma_0}\right)^2\right], \qquad \frac{\partial p(x,t=0,x_0)}{\partial t} = 0, \label{initial_cond}
\end{equation}
with $\sigma_0$ being the width of the pulse determining the frequencies to span.

\subsection{Boundary conditions}
We consider impedance boundaries and denote the boundary domain as $\Gamma$ (in 1D, the left and right endpoints). We will omit the source position $x_0$ in the following.
\subsubsection{Frequency-independent impedance boundaries}
The acoustic properties of a wall can be described by its surface impedance\cite{kuttruff2016room} $Z_s = \frac{p}{v_n}$ where $v_n$ is the normal component of the velocity at the same location on the wall surface. Combining the surface impedance with the pressure term $\frac{\partial p}{\partial \mathbf{n}} = -\rho_0 \frac{\partial v_n}{\partial t}$ of the linear coupled wave equation yields
\begin{equation}
	\frac{\partial p}{\partial t} = -c\xi\frac{\partial p}{\partial \mathbf{n}}, \label{impedance_bound_cond}
\end{equation}
where $\xi = Z_s/(\rho_0c)$ is the normalized surface impedance and $\rho_0$ denotes the air density ($\text{kg}/\text{m}^3$). Note that perfectly reflecting boundaries can be obtained by letting $\xi \rightarrow \infty$ being the Neumann boundary formulation.

\subsubsection{Frequency-dependent impedance boundaries}
The wall impedance can be written as a rational function in terms of the admittance $Y = 1/Z_s$ and rewritten by using partial fraction decomposition in the last equation \cite{Troian2017}
\begin{equation}
  Y(\omega) = \frac{a_0 + \ldots + a_N(-i\omega)^N}{1+\ldots+b_N(-i\omega)^N} = Y_{\infty} + \sum_{k=0}^{Q-1}\frac{A_k}{\lambda_k - i\omega} + \sum_{k=0}^{S-1}\left( \frac{B_k + iC_k}{\alpha_k + i\beta_k - i\omega} + \frac{B_k - iC_k}{\alpha_k -i\beta_k -i\omega} \right), \label{rational_func}
\end{equation}
where $a,b$ are real coefficients, $i = \sqrt{-1}$ being the complex number, $Q$ is the number of real poles $\lambda_k$, $S$ is the number of complex conjugate pole pairs $\alpha_k \pm j\beta_k$, and $Y_{\infty}$, $A_k$, $B_k$ and $C_k$ are numerical coefficients. Since we are concerned with the (time-domain) wave equation, the inverse Fourier transform is applied on the admittance and on the partial fraction decomposition term in \autoref{rational_func}. Combining these gives \cite{Troian2017}
\begin{align}
  v_n(t) = Y_{\infty}p(t) + \sum_{k=0}^{Q-1}A_k\phi_k(t) + \sum_{k=0}^{S-1}2\left[ B_k \psi_k^{(0)}(t) + C_k\psi_k^{(1)}(t) \right]. \label{velocity_bound_freq_dep}
\end{align}
The functions $\phi_k$, $\psi_k^{(0)}$, and $\psi_k^{(1)}$ are the so-called accumulators determined by the following set of ordinary differential equations (ODEs) referred to as auxiliary differential equations (ADEs)
\begin{equation}
  \frac{d\phi_k}{dt} + \lambda_k\phi_k = p, \qquad
  \frac{d\psi_k^{(0)}}{dt} + \alpha_k\psi_k^{(0)} + \beta_k\psi_k^{(1)} = p, \qquad
  \frac{d\psi_k^{(1)}}{dt} + \alpha_k\psi_k^{(1)} - \beta_k\psi_k^{(0)} = 0. \label{ade_diff_eq}
\end{equation}
The boundary conditions can then be formulated by inserting the velocity $v_n$ calculated in \autoref{velocity_bound_freq_dep} into the pressure term of the linear coupled wave equation $\frac{\partial p}{\partial \mathbf{n}} = -\rho_0 \frac{\partial v_n}{\partial t}$. 
\subsection{PINNs}
\begin{figure*}[ht]
  \centerline{\includegraphics[width=0.95\textwidth]{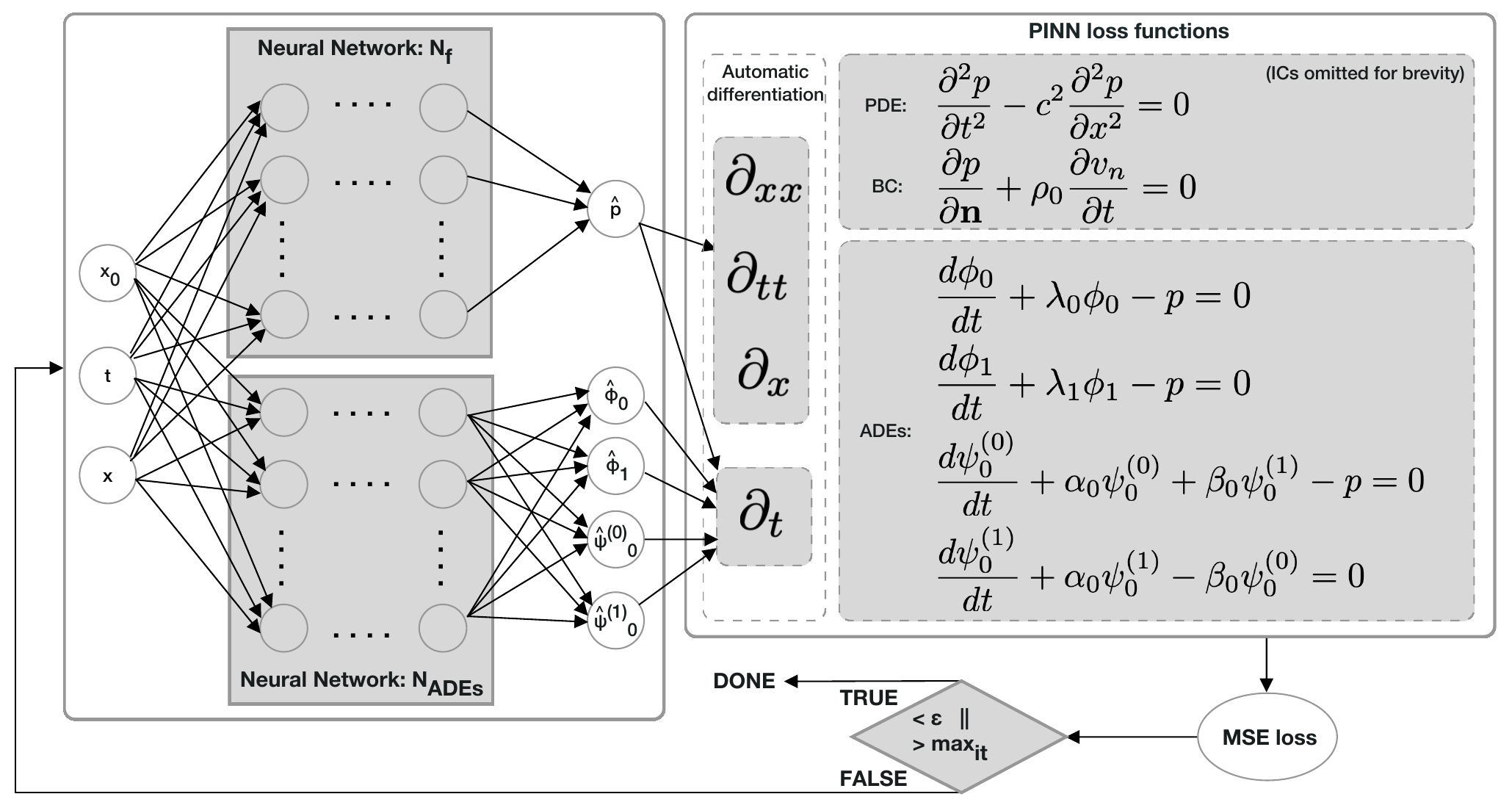}}
  \caption{\label{fig:nn_freq_dep_pinn}PINN scheme for frequency-dependent boundaries. Left: Two fully connected Feed-Forward Neural Network architectures, $\mathcal{N}_f$ (PDE + ICs) and $\mathcal{N}_\text{ADE}$ (ADEs). Right: The governing physical equations and ADEs are coupled via the loss function (ICs and scaling terms are omitted for brevity). Training is done when a maximum number of epochs is reached, or the total loss is smaller than a given threshold.}
\end{figure*}

Two multi-layer feed-forward neural networks are setup
\begin{equation*}
  \hat{f}: (x,t,x_0) \mapsto \mathcal{N}_f(x,t,x_0; \mathbf{W}, \mathbf{b}), \qquad \hat{g}: (x,t,x_0) \mapsto \mathcal{N}_\text{ADE}(x,t,x_0; \mathbf{W}, \mathbf{b}),
\end{equation*}
where $\mathbf{W}$ and $\mathbf{b}$ are the network weights and biases, respectively; $\mathcal{N}_\text{ADE}$ is only applied in case of frequency-dependent boundaries.
The networks take three inputs $x,t,x_0$ corresponding to the spatial, temporal, and source position dimensions. The network $\mathcal{N}_f$ has one output $\hat{p}(x,t,x_0)$ approximating $p(x,t,x_0)$ predicting the pressure; the network $\mathcal{N}_\text{ADE}$ is a multi-output network with the number of outputs corresponding to the number of accumulators to approximate as explained in the following.

Including only information about the underlying physics, the governing partial differential equation (PDE) from \autoref{wave_equation} and initial conditions (ICs) from \autoref{initial_cond} can be learned by minimizing the mean squared error loss denoted $\Vert \bullet \Vert$ as
\begin{equation}
  \arg \min_{\mathbf{W,b}} \loss(\mathbf{W}, \mathbf{b}) = \loss_\text{PDE} + \lambda_\text{IC}\loss_\text{IC} + \lambda_\text{BC}\loss_\text{BC} + \loss_\text{ADE}, \label{loss_terms}
\end{equation}
where
\begin{subequations}
\begin{eqnarray}
  \loss_\text{PDE} &=& \Vert \frac{\partial^2}{\partial t^2} \mathcal{N}_f(\xf, \tf, \srcf; \mathbf{W}, \mathbf{b}) -c^2\nabla^2 \mathcal{N}_f(\xf, \tf, \srcf; \mathbf{W}, \mathbf{b}) \Vert \label{pde_loss}, \\
  \loss_\text{IC} &=& \Vert \mathcal{N}_f(\xic, 0, \sic; \mathbf{W}, \mathbf{b}) - \exp\left[-\left(\frac{\xic-\sic}{\sigma_0}\right)^2\right] \Vert + \Vert \frac{\partial}{\partial t} \mathcal{N}_f(\xic, 0, \sic; \mathbf{W}, \mathbf{b}) \Vert. \label{ic_loss}
\end{eqnarray}
\end{subequations}
Here, $\{ \xic,\sic \}_{i=1}^{N_\text{IC}}$ denotes the initial $N_\text{IC}$ data points, $\{ \xf,\tf,\srcf \}_{i=1}^{N_{f}}$ denotes the $N_{f}$ collocation points for the PDE $f$, and the penalty weights $\lambda_\text{IC}$ and $\lambda_\text{BC}$ are used for balancing the impact of the individual terms.  The loss function $\loss_\text{BC}$ will be treated separately for the impedance boundary conditions in the following, where $\{ \xbc,\tbc,\sbc \}_{i=1}^{N_\text{BC}}$ will, correspondingly, be denoting the $N_\text{BC}$ collocation points on the boundaries. For frequency-dependent boundaries, an auxiliary neural network will be coupled, resulting in the additional loss term $\loss_\text{ADE}$ in \autoref{loss_terms} and is explained in detail in the following.

\subsubsection{Frequency-independent impedance boundary loss functions}
The frequency-independent boundary condition, \autoref{impedance_bound_cond}, is included in the loss function $\loss_\text{BC} := \loss_\text{INDEP}$ satisfied by $\loss_\text{INDEP} = \Vert \frac{\partial}{\partial t} \mathcal{N}_f(\xbc, \tbc, \sbc; \mathbf{W}, \mathbf{b}) + c\xi\frac{\partial}{\partial \mathbf{n}}\mathcal{N}_f(\xbc, \tbc, \sbc; \mathbf{W}, \mathbf{b}) \Vert$.

\subsubsection{Frequency-dependent impedance boundary loss functions}
For frequency-dependent boundaries, the ADEs need to be solved as well, approximating the ODEs from \autoref{ade_diff_eq} by introducing an additional neural network $\mathcal{N}_\text{ADE}(x_b,t,x_0; \mathbf{W}, \mathbf{b})$ parameterized by $x_b$ and $x_0$ for boundary positions and moving sources, respectively. The network has multiple outputs $\mathcal{N}_\text{ADE}(x_b,t,x_0; \mathbf{W}, \mathbf{b}) = [\phit_0, \phit_{1}, \ldots, \phit_{Q-1}; \psit^{(0)}_0, \psit^{(0)}_1, \ldots, \psit^{(0)}_{S-1}; \psit^{(1)}_0, \psit^{(1)}_1, \ldots, \psit^{(1)}_{S-1}]$ corresponding to the {\it scaled} accumulators determined by a scaling factor $l_\text{ADE}^{\bullet}$, mapping $\phit_k = l_\text{ADE}^{\phi_k}\hat{\phi}_k$, $\psit^{(0)}_k = l_\text{ADE}^{\psi_k}\hat{\psi}^{(0)}_k$ and $\psit^{(1)}_k = l_\text{ADE}^{\psi_k}\hat{\psi}^{(1)}_k$, such that $\phit_k, \psit^{(0)}_k, \psit^{(1)}_k: (x_b,t,x_0) \mapsto [-1,1]$ matching the range of the tanh function used in this work. The scaling factors are independent of the geometry and domain dimensionality, only the material properties determine the amplitude of the accumulators. In this work, the scaling factors are determined using the SEM solver, but might be analytically estimated from the accumulators considering only a single reflection in a 1D domain. A graphical representation of the neural network architectures for approximating the governing physical equations and ADEs is depicted in \autoref{fig:nn_freq_dep_pinn}. 
The accumulators with parametrized moving sources and boundary positions $\phit_k(\xbc,\tbc,\sbc)$, $\psit^{(0)}_k(\xbc,\tbc,\sbc)$, $\psit^{(1)}_k(\xbc,\tbc,\sbc)$ can be learned by minimizing the mean squared error loss as (arguments omitted)
\begin{equation}
  \arg \min_{\mathbf{W,b}} \loss_\text{ADE}(\mathbf{W}, \mathbf{b}) = \sum_k^{Q-1}\tilde{\lambda}_\text{ADE}^{\phit_k}\loss_{\phit_k} + \sum_k^{S-1}\left(\tilde{\lambda}_\text{ADE}^{\psit_k^{(0)}}\loss_{\psit^{(0)}_k} + \tilde{\lambda}_\text{ADE}^{\psit_k^{(1)}}\loss_{\psit^{(1)}_k}\right), \label{loss_terms_accum}
\end{equation}
where
\begin{subequations}
\begin{eqnarray}
  \loss_{\phit_k} &=& \Vert \frac{\partial}{\partial t}\phit_k +\lambda_k\phit_k - l_\text{ADE}^{\phi_k}\mathcal{N}_f \Vert, \\
  \loss_{\psit^{(0)}_k} &=& \Vert \frac{\partial}{\partial t}\psit^{(0)}_k + \alpha_k\psit^{(0)}_k + \beta_k l_\text{ADE}^{\psi_k^{(0)}} (1/l_\text{ADE}^{\psi_k^{(1)}}) \psit^{(1)}_k - l_\text{ADE}^{\psi_k^{(0)}}\mathcal{N}_f \Vert, \\
  \loss_{\psit^{(1)}_k} &=& \Vert \frac{\partial}{\partial t}\psit^{(1)}_k + \alpha_k\psit^{(1)}_k - \beta_k l_\text{ADE}^{\psi_k^{(1)}} (1/l_\text{ADE}^{\psi_k^{(0)}}) \psit^{(0)}_k \Vert,
\end{eqnarray}
\end{subequations}
and
\begin{equation*}
  \tilde{\lambda}_\text{ADE}^{\phit_k} = (1/l_\text{ADE}^{\phi_k}) \lambda_\text{ADE}^{\phit_k}, \qquad \tilde{\lambda}_\text{ADE}^{\psit_k^{(0)}} = (1/l_\text{ADE}^{\psi_k^{(0)}})\lambda_\text{ADE}^{\psit_k^{(0)}}, \qquad \tilde{\lambda}_\text{ADE}^{\psit_k^{(1)}} = (1/l_\text{ADE}^{\psi_k^{(1)}})\lambda_\text{ADE}^{\psit_k^{(1)}}.
\end{equation*}
The frequency-dependent boundary conditions are satisfied by $\loss_\text{DEP} = \Vert \frac{\partial}{\partial \mathbf{n}} \mathcal{N}_f(\xbc, \tbc, \sbc; \mathbf{W}, \mathbf{b}) + \rho_0 \frac{\partial v_{n}(\xbc,\tbc,\sbc)}{\partial t} \Vert$, where $v_{n}$ is the expression at the boundaries given in \autoref{velocity_bound_freq_dep} with $\hat{\phi}_k = 1/l^{\phi_k}_\text{ADE}\phit_k$, $\hat{\psi}^{(0)}_k = 1/l^{\psi_k^{(0)}}_\text{ADE}\psit^{(0)}_k$ and $\hat{\psi}^{(1)}_k = 1/l^{\psi_k^{(1)}}_\text{ADE}\psit^{(1)}_k$, ensuring that the accumulators are properly re-scaled. The loss is included as the term $\loss_\text{BC} := \loss_\text{DEP}$ in \autoref{loss_terms} together with the loss for the ADEs in \autoref{loss_terms_accum}.

\subsection{Setup}\label{sec:setup}
TensorFlow 2.5.1 \cite{tensorflow}, SciANN 0.6.4.7 \cite{Haghighat2021}, and Python 3.8.9 with 64 bit floating points for the Neural Network weights are used. The code for reproducing the results can be found online: \url{https://github.com/dtu-act/pinn-acoustic-wave-prop}.
 
Reference data for impedance boundaries are generated using a fourth-order Jacobi polynomial SEM solver. The grid was discretized with 20 points per wavelength spanning frequencies up to 1000 Hz yielding an average grid resolution of $\Delta x = 0.017$ m, and the time step was $\Delta t=\text{CFL}\times\Delta x/c$, where CFL is the Courant-Friedrichs-Lewy constant ($\text{CFL} = 1.0$ and $\text{CFL} = 0.1$ for frequency-independent and dependent boundaries, respectively). The speed of sound $c=1$ m/s is used for the PINN setup, implying $\Delta x = \Delta t/c = \Delta t$, which is a normalization introduced to ensure the same scaling in time and space required for the optimization problem to converge. In case of a normalized speed of sound, the effective normalized frequency is correspondingly $f = f_\text{phys}/c_\text{phys}$, since the wave is now travelling slower compared to the physical setup. To evaluate the results for a physical speed of sound $c_\text{phys} = 343$ m/s, the temporal dimension should be converted back as $t_\text{phys} = t/c_\text{phys}$ sec. In case of frequency-dependent boundaries, the velocity from \autoref{velocity_bound_freq_dep} needs to be normalized accordingly regarding frequency and flow resistivity $\sigma_\text{mat} = \sigma_\text{mat, phys}/c_\text{phys}$ in Miki's model. Fitting the parameters for $c=1$ m/s yields modified $\lambda_k$, $\alpha_k$, $\beta_k$ and $Y_{\infty}$ values resulting in the exact same surface impedance as for $c=343$ m/s, but scaled by $c_\text{phys}$ in frequency range and amplitude. This can be seen from the complex wavenumber and characteristic impedance of the porous medium in Miki's model involving $\frac{f}{\sigma_\text{mat}}$ and $\frac{2\pi f}{c}$ not being affected by normalization\footnote{See
Appendix for addition results for Neumann boundaries, accumulator predictions, runtime efficiency of the surrogate model, and detailed explanation of the normalization for the frequency-dependent impedance boundary formulation.}.

The point distribution in time and space, number of sources, penalty weights $\lambda$ and scaling factors $l_\text{ADE}$ for the ADEs are listed in \autoref{tab:network_params}. Note that the number of (time and space) domain points (30\%) per source (7) is $47,089\cdot 0.3/7 = 2018$, satisfying the Nyquist theorem, since $\Delta x = 2/\sqrt{2018} = 0.045$ m (we can use the square root to get the gridpoint distribution in the spatial dimension) resulting in $\text{ppw} = \lambda_w/\Delta x = 7.6 $ points per wavelength; $\lambda_w=c/f$ being the wavelength for physical frequency 1000 Hz and physical speed of sound 343 m/s. For the neural network, we have used the ADAM optimizer and the mean-squared error for calculating the losses for both networks. The training was run with learning rate 1e-4 and batch size 512 until a total loss of $\epsilon=\text{2e-4}$ was reached (roughly 16k and 20k epochs needed for frequency-independent and dependent boundaries, respectively). The relatively big batch size was chosen to ensure that enough initial and boundary points were included in the optimization steps.

The network architecture of $\mathcal{N}_f$ consists of three layers, each with 256 neurons applying the sine activation function in each layer except for a linear output layer, with proper weight initialization\cite{sitzmann2020implicit}. Using sine activation functions can be seen as representing the signal using Fourier series \cite{benbarka2021seeing} and is probably the reason for a significantly better convergence compared to using the more common choice of tanh activation functions. However, experiments showed degraded interpolation properties using sine activation functions when the network was trained on grids with source positions distributed more sparsely (0.3 m), even when lowering the number of neurons to prevent overfitting. A reason could be related to the distributed source interval violating the Nyquist sampling theorem $\Delta x < c/(2f) = 0.17$ m and causing aliasing effects, but this remains an open question. Therefore, the source positions were distributed evenly with finer resolution to improve the results between the source positions, consequently resulting in a sparser grid per source by keeping the total number of points the same. Despite the sparser grid, the convergence and final error still showed satisfying results. Distributing the source positions more densely is trivial in a data-free implementation, but if a combination of the underlying physics and simulated/measured data is considered later, a large number of source positions could be practically challenging.

The network architecture of $\mathcal{N}_\text{ADE}$ consists of three layers, each with 20 neurons applying the tanh activation function in each layer except for a linear output layer, with Glorot normal initialization of the weights  \cite{pmlr-v9-glorot10a}. Using the tanh function is an obvious choice since we have chosen to scale the accumulators to take values in the range $[-1,1]$.

\begin{table}
  \centering
  \begin{tabular}{ccccc|ccccccc}
    \hline\hline
    \#total & \#BC & \#IC & \#inner\footnotemark[1] & \#srcs & $\lambda_\text{IC}$ & $\lambda_\text{BC}$ & $\lambda_\text{ADE}^{\bullet}$ & $l_\text{ADE}^{\phi_0}$ & $l_\text{ADE}^{\phi_1}$ & $l_\text{ADE}^{\psi^{(0)}_0}$ & $l_\text{ADE}^{\psi^{(1)}_0}$ \\\hline
    $47,089$ & 45\% & 25\% & 30\% & 7 & 20 & 1 & 10 & 10.3 & 261.4 & 45.9 & 22.0 \\
    \hline\hline
  \end{tabular}
  \vspace{0.2cm}
  \caption{\label{tab:network_params} Number of points in time and space; inner domain, boundaries, and initial condition point distributions; number of evenly distributed sources (srcs); values for the penalty weights $\lambda$; scaling factors $l$ for normalizing the accumulators.}
  \footnotetext[1]{The centered Latin Hypercube Sampling strategy \cite{Stein1987} is used.}  
\end{table}

\section{Results}\label{sec:results}
Frequency-independent and dependent boundary conditions are tested, each with parameterized moving sources trained at seven evenly distributed positions $\mathbf{x}_0 = [-0.3,-0.2,\ldots,0.3]$ m and evaluated at five positions $\mathbf{x}_0 = [-0.3,-0.15,0.0,0.15,0.3]$ m. Additional results are included in the Appendix\footnotemark[\value{footnote}]. The source is satisfied through the initial condition modeled as a Gaussian impulse from \autoref{initial_cond} with $\sigma_0=0.2$ spanning frequencies up to $1000$Hz. The speed of sound $c_\text{phys}=343$ m/s and air density $\rho_0 = 1.2 \text{ kg}/\text{m}^3$ are used for all studies.

\begin{figure*}
  \begin{subfigure}{0.95\textwidth}
    \includegraphics[width=1.0\textwidth]{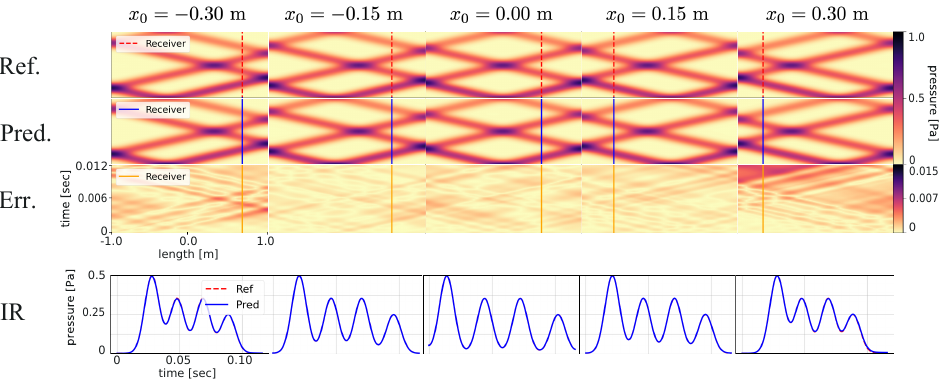}
    \subcaption{Frequency-independent boundaries.}\label{fig:freq_indep_results}
    \includegraphics[width=1.0\textwidth]{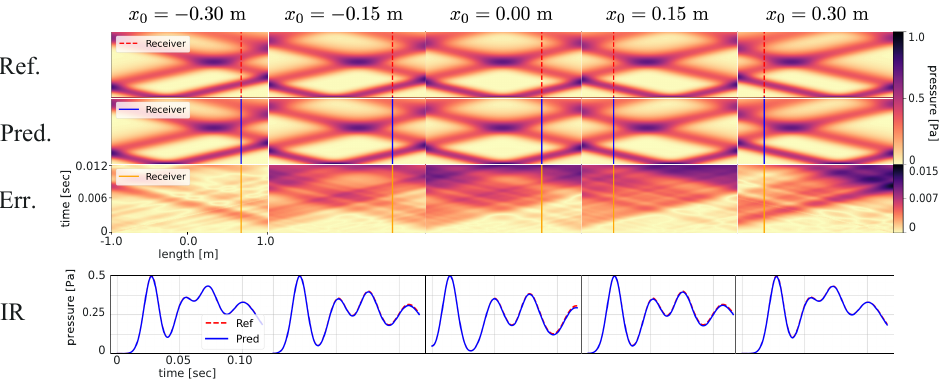}
    \subcaption{Frequency-dependent boundaries.}\label{fig:freq_dep_results}
  \end{subfigure}
  \caption{Wave propagations in a 1D domain $[-1,1]$ m.}
\end{figure*}
First, we test the frequency-independent boundary condition with normalized impedance $\xi=5.83$ depicted in \autoref{fig:freq_indep_results}. Then, we test the frequency-dependent impedance boundary condition, where the boundary is modeled as a porous material mounted on a rigid backing with thickness $d_\text{mat}=0.10$ m with an air flow resistivity of $\sigma_\text{mat,phys} = 8,000 \text{ Nsm}^{-4}$. The surface impedance $Y$ of this material is estimated using Miki's model \cite{Miki199019} and mapped to a two-pole rational function in the form of \autoref{rational_func} with $Q=2$ and $S=1$ using a vector fitting algorithm \cite{Gustavsen1999} yielding the coefficients for \autoref{velocity_bound_freq_dep}. The results are depicted in \autoref{fig:freq_dep_results}.

We observe that the shape of the wave propagations is well captured, and the impulse responses also fit the reference solutions very well for all boundary types. The relative mean error $\mu_\text{rel}(x,x_0) = \frac{1}{N}\sum_{i=0}^{N-1} \frac{\vert \hat{p}(x,t^i,x_0) - p(x,t^i,x_0)\vert}{p(x,t^i,x_0)}$ within $-60$dB and absolute maximum error $\infty_\text{abs}(x,x_0) = \max\{\vert \hat{p}(x,t^i,x_0) - p(x,t^i,x_0)\vert\ : i=0\ldots N-1\}$ of the impulse responses originating from various source and receiver positions are summarized in \autoref{tab:pred_errors}. Relative errors are below 2\%/0.2 dB for all predictions. The absolute maximum errors are below 0.011 Pa for all predictions indicating that no severe outliers are present.
\begin{table}
  \centering  
  \begin{tabular}{l|cc|cc|cc|cc|cc}
    \hline\hline
    & \multicolumn{2}{c|}{$s_0$} & \multicolumn{2}{c|}{$s_1$} & \multicolumn{2}{c|}{$s_2$} & \multicolumn{2}{c|}{$s_3$} & \multicolumn{2}{c}{$s_4$} \\
    & $\mu_\text{rel}$ & $\infty_\text{abs}$ & $\mu_\text{rel}$ & $\infty_\text{abs}$ & $\mu_\text{rel}$ & $\infty_\text{abs}$ & $\mu_\text{rel}$ & $\infty_\text{abs}$ & $\mu_\text{rel}$ & $\infty_\text{abs}$ \\\hline
    \bf{Freq. indep.} & 
    0.7\% & 0.004 &
    0.3\% & 0.002 &
    0.5\% & 0.001 &
    0.4\% & 0.002 &
    1.5\% & 0.006 \\
    \bf{Freq. dep.} & 
    0.5\% & 0.004 &
    1.5\% & 0.008 &
    1.7\% & 0.011 &
    1.9\% & 0.009 &
    1.1\% & 0.005 \\
    \hline\hline
  \end{tabular}
  \vspace{0.2cm}
  \caption{\label{tab:pred_errors} Time domain errors for source/receiver pairs $(x_0,x)$ measured in meters in what follows at positions $s_0 = (-0.3,0.64)$, $s_1 = (-0.15,0.58)$, $s_2 = (0.0,0.5)$, $s_3 = (0.15,-0.58)$, $s_4 = (0.3,-0.66)$. $\mu_\text{rel}(x,x_0) = \frac{1}{N}\sum_{i=0}^{N-1} \frac{\vert \hat{p}(x,t^i,x_0) - p(x,t^i,x_0)\vert}{p(x,t^i,x_0)}$ is the relative mean error over time within $-60$dB range and $\infty_\text{abs}(x,x_0) = \max\{\vert \hat{p}(x,t^i,x_0) - p(x,t^i,x_0)\vert\ : i=0\ldots N-1\}$ Pa is the maximum absolute error for source/receiver pair $s_i$.}  
\end{table}

\section{Conclusion and future work}
A novel method is presented for predicting the sound field in a 1D domain for impedance boundaries and parameterized moving Gaussian sources using PINNs. A coupled system of differential equations, consisting of the governing physical equations and a system of ODEs predicting the accumulators of the ADEs, was used for training the PINN. The equations for the ADEs depend only on time $t$ but were parameterized to take boundary and source positions into account, yielding a very flexible implementation. The results are promising, with relative mean errors below 2\%/0.2 dB for all cases. The approach taken by learning a compact surrogate model that is inexpensive to evaluate at runtime shows potential to overcome current numerical methods' limitations in modeling flexible scenes, such as moving sources.

Compared to standard numerical methods, the PINN method takes up to three orders of magnitude more time to converge. Therefore, to solve realistic problems in 3D, the convergence rate needs to be improved. This is partly due to the need for fairly large amounts of grid points with 70\% of the points located at the initial time step and at boundaries where also penalty weights are needed for balancing each loss term. Formulating an ansatz imposing initial and boundary conditions directly could overcome this problem\cite{Sukumar2021}. Also, considering other architectures taking (discrete) time-dependence into account instead of optimizing the entire spatio-temporal domain at once might improve the learning rate and produce more precise results. Moreover, we have observed challenges in the global optimizer for a larger domain size and/or by increasing the frequency due to the ratio between zero and non-zero pressure values. Domain decomposition methods \cite{Shukla2021} have been introduced to overcome this limitation. In ongoing work, more complex benchmarks are being considered.

\section{Acknowledgments}
Thanks to DTU Computing Center GPULAB for access to GPU clusters and swift help. Also, a big thanks to Ehsan Haghighat for valuable discussions regarding PINNs and SciANN. Last but not least, thanks to Finnur Pind for making an SEM code available for calculating reference solutions.

\bibliography{./pinns}{}
\bibliographystyle{plain}

\section{Appendix}
We provide here additional results for wave propagation predictions with Neumann boundary conditions, accumulator predictions, and runtime efficiency using PINNs.
\subsection{Methods}
Analytical solutions are used for Neumann boundary conditions to assess the quality of the developed PINN models. The Neumann boundary condition is given as
\begin{align}
  \frac{\partial p(x,t)}{\partial \mathbf{n}} = 0 \label{neumann_bound_cond}, \qquad x \in \Gamma,
\end{align}
where $\mathbf{n}$ is the normal pointing outwards from the boundary $\Gamma$. This condition corresponds to a `hard wall' allowing displacement of the particles in the medium with the net directional force to be zero. Neumann boundary condition \autoref{neumann_bound_cond} is satisfied by
\begin{equation}  
  \loss_{neumann} = \left \Vert \frac{\partial}{\partial \mathbf{n}} \mathcal{N}_f(\xbc, \tbc, \sbc; \mathbf{W}, \mathbf{b}) \right \Vert,
\end{equation}
and included as the term for handling the boundary condition $\loss_{BC} := \loss_{neumann}$.
\subsubsection*{Governing equations}
We consider in the following the use of PINNs for the construction of a surrogate model predicting the solution to the linear acoustic wave equation in 1D
\begin{equation}
  \frac{\partial^2 p(x,t)}{\partial t^2}-c^2\frac{\partial^2 p(x,t)}{\partial x^2}=0,\qquad t \in \mathbb{R^+}, \qquad x \in \mathbb{R}. \label{wave_equation}
\end{equation}
where $p$ is the pressure (Pa), $t$ is the time (s) and $c$ is the speed of sound (m/s). An analytical solution with source position $x_0 \in \mathbb{R}$ is 
\begin{equation}
p(x,t,x_0) = \frac{1}{2}\exp\left[-\left(\frac{x-x_0 - ct}{\sigma_0}\right)^2\right] + \frac{1}{2}\exp\left[-\left(\frac{x-x_0 + ct}{\sigma_0}\right)^2\right], \label{analytical_sol}
\end{equation}
with initial conditions (ICs)
\begin{equation}
p(x,t=0,x_0) = \exp\left[-\left(\frac{x-x_0}{\sigma_0}\right)^2\right], \qquad \frac{\partial p(x,t=0,x_0)}{\partial t} = 0, \label{initial_cond}
\end{equation}
effectively being a Gaussian impulse with $\sigma_0$ being the width of the pulse determining the frequencies to span.

\subsubsection*{Implications for Miki's model for normalized speed of sound}
Miki's model\cite{Miki199019} can be used to estimate the frequency-dependent boundary condition of a porous material mounted on a rigid backing with thickness $d_\text{mat}$ m with the flow resistivity $\sigma_\text{mat}$\cite{Allard2009}
\begin{equation}
  Z_s(f) = -iZ_c\cot(k_t(f) \cdot d_\text{mat}),
\end{equation}
where
\begin{subequations}
\begin{eqnarray}
  Z_c(f) &=& \rho_0 c\left(1+0.07\left(\frac{f}{\sigma_\text{mat}}\right)^{-0.632} - i0.107\left(\frac{f}{\sigma_\text{mat}}\right)^{-0.632}\right), \\
  k_t(f) &=& \frac{2\pi f}{c}(1+0.109\left(\frac{f}{\sigma_\text{mat}}\right)^{-0.618} - i0.16\left(\frac{f}{\sigma_\text{mat}})^{-0.618}\right),
\end{eqnarray}
\end{subequations}
$i = \sqrt{-1}$ being the complex number, and $\rho_0$ the air density. We can quickly convince ourselves that the terms $\frac{f}{\sigma_\text{mat}}$ and $\frac{2\pi f}{c}$ yield the same results no matter the frequency and flow resistivity are both normalized or not. This means that the normalized impedance corresponds precisely to the physical impedance but scaled by $c_\text{phys}$ in frequency range and amplitude -- the latter due to the term $\rho_0 c$ in $Z_c$. Therefore, by normalizing the frequency and flow resistivity, we ensure that the boundary condition of the porous layer remains the same as in the physical (un-normalized) setup.

\subsection{Results}
\begin{figure*}
  \includegraphics[width=1.0\textwidth]{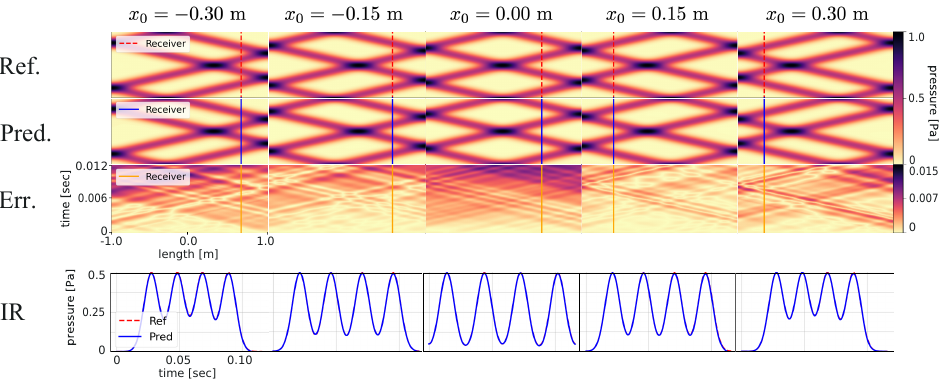}
\caption{Wave propgations in a 1D domain $[-1,1]$ m for Neumann boundary conditions.}\label{fig:neumann_results}
\end{figure*}

\begin{figure*}
  \includegraphics[width=1.0\textwidth]{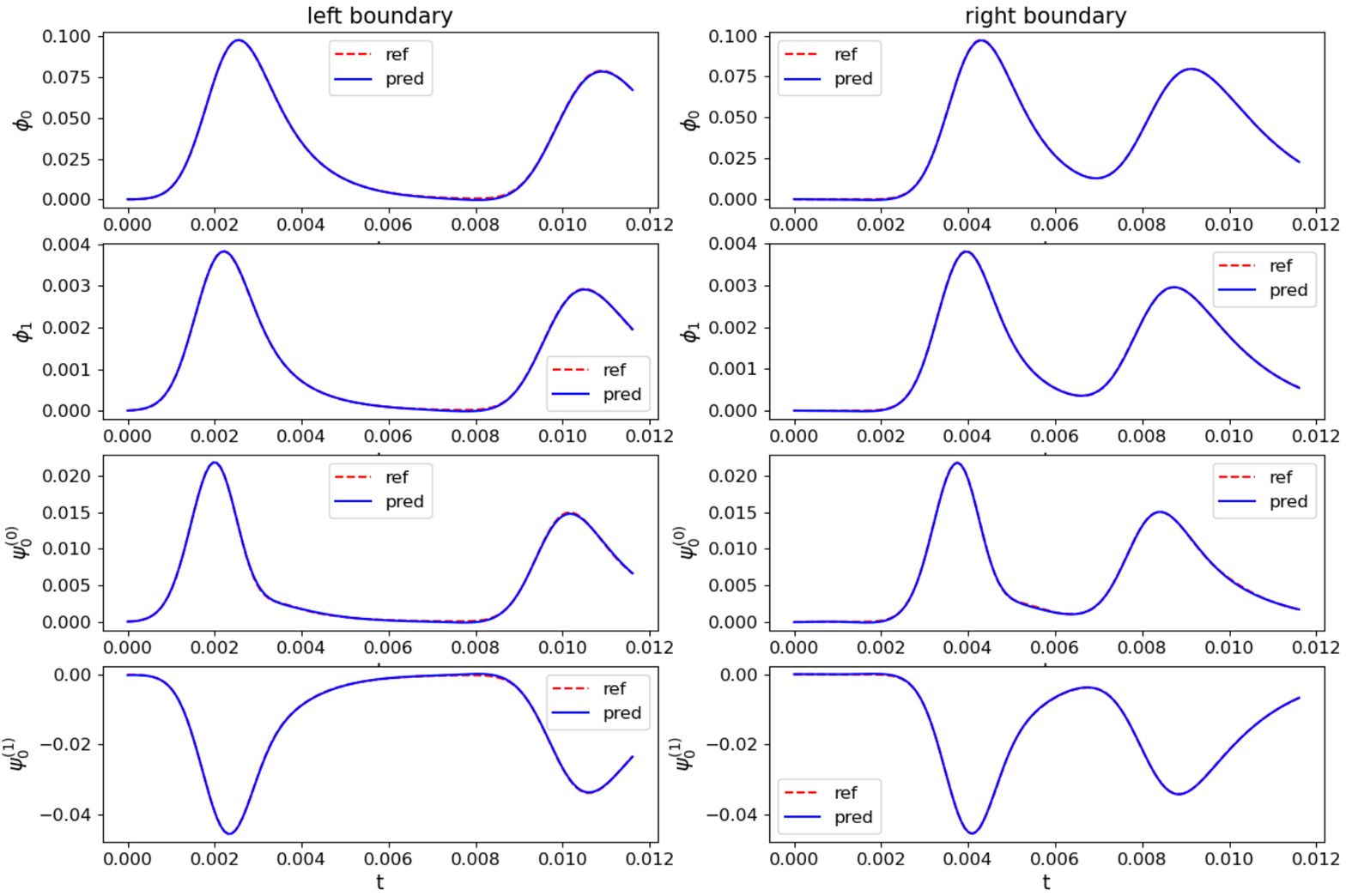}
  \caption{Predicted accumulators compared with reference solutions for $x_0=-0.3$ m.}\label{fig:accumulator_results}
\end{figure*}

Reference data is calculated from the analytical solution given in \autoref{analytical_sol}. First, Neumann boundary conditions is tested and depicted in \autoref{fig:neumann_results}, each with parameterized moving sources trained at seven evenly distributed positions $\mathbf{x}_0 = [-0.3,0.2,\ldots,0.3]$ m and evaluated at five positions $\mathbf{x}_0 = [-0.3,-0.15,0.0,0.15,0.3]$ m. The source is satisfied through the initial condition modeled as a Gaussian impulse from \autoref{initial_cond} with $\sigma_0=0.2$ spanning frequencies up to $1000$ Hz.
We observe that the shape of the wave propagations is well captured, and the impulse responses also fit the reference solutions very well. The relative mean error $\mu_\text{rel}(x,x_0) = \frac{1}{N}\sum_{i=0}^{N-1} \frac{\vert \hat{p}(x,t^i,x_0) - p(x,t^i,x_0)\vert}{p(x,t^i,x_0)}$ within $-60$dB and absolute maximum error $\infty_\text{abs}(x,x_0) = \max\{\vert \hat{p}(x,t^i,x_0) - p(x,t^i,x_0)\vert\ : i=0\ldots N-1\}$ of the impulse responses originating from various source and receiver positions are summarized in \autoref{tab:pred_neumann_errors}. Relative errors are below 3\%/0.3 dB for all predictions. The absolute maximum errors are below 0.011 Pa for all predictions indicating that no severe outliers are present.
\begin{table}[h!]
  \centering  
  \begin{tabular}{l|cc|cc|cc|cc|cc}
    \hline\hline
    & \multicolumn{2}{c|}{$s_0$} & \multicolumn{2}{c|}{$s_1$} & \multicolumn{2}{c|}{$s_2$} & \multicolumn{2}{c|}{$s_3$} & \multicolumn{2}{c}{$s_4$} \\
    & $\mu_{rel}$ & $\infty_{abs}$ & $\mu_{rel}$ & $\infty_{abs}$ & $\mu_{rel}$ & $\infty_{abs}$ & $\mu_{rel}$ & $\infty_{abs}$ & $\mu_{rel}$ & $\infty_{abs}$ \\\hline
    \bf{Neumann} & 
    1.3\% & 0.011 &
    0.8\% & 0.008 &
    2.8\% & 0.01 &
    0.6\% & 0.005 &
    0.7\% & 0.005 \\
    \hline\hline
  \end{tabular}
  \vspace{0.2cm}
  \caption{\label{tab:pred_neumann_errors} Time domain errors for source/receiver pairs $(x_0,x)$ measured in meters in what follows at positions $s_0 = (-0.3,0.64)$, $s_1 = (-0.15,0.58)$, $s_2 = (0.0,0.5)$, $s_3 = (0.15,-0.58)$, $s_4 = (0.3,-0.66)$. $\mu_{rel} = \frac{\vert p_{pred} - p_{ref}\vert}{p_{ref}}$ is the relative mean error over time within $-60$dB range and $\infty_{abs} = \max(\vert p_{pred} - p_{ref}\vert)$ Pa is the maximum absolute error for source/receiver pair $s_i$.}
\end{table}
Secondly, we show the predictions for the accumulators for frequency-dependent boundaries in \autoref{fig:accumulator_results}. We see that the predicted accumulators fit the references very well.

Lastly, we assess the runtime efficiency of the surrogate model. An impulse response (IR) of length $t=0.5$ sec sampled at 44.1 kHz is retrieved at receiver position $x_\text{receiver}=0.5$ m for source position $x_0 = 0.0$ m. The timings for the surrogate models are all below 64 ms (using Nvidia Tesla V100 16 GB graphics card) for evaluating a single IR consisting of 22,050 samples. The execution time of our surrogate model can be considered real-time\cite{sandvad1996dynamic}, where experiments showed that a latency above 96 ms would introduce significant degrading effects on the azimuth error and elapsed time (timing misjudgments). Note that bigger and more complicated neural networks might be needed for realistic 3D scenes, being more expensive to execute and may require implementation techniques to optimize the execution\cite{Queiruga2020}.

\end{document}